\documentclass[sigconf,natbib=true,noacm]{acmart}
\usepackage{booktabs}
\usepackage{multirow}
\usepackage{xcolor}
\usepackage{colortbl}
\usepackage{array}
\usepackage{multirow}
\usepackage{makecell}
\usepackage{algorithm}
\usepackage{algorithmic}
\AtBeginDocument{%
  }

\setcopyright{acmlicensed}
\copyrightyear{2026}  
\acmYear{2026}        
\acmDOI{XXXXXXX.XXXXXXX}  

\acmConference[Conference acronym 'XX]{Full Conference Name}{Month DD--DD,
  2025}{City, Country}  

\acmISBN{978-1-4503-XXXX-X/2025/06}  

\begin{document}

\title{When Vision Meets Texts in Listwise Reranking}

\author{Hongyi Cai}
\email{xcloudfance@gmail.com}  
\orcid{0000-0000-0000-0000}   
\affiliation{%
  \institution{Universiti Malaya}
  \city{Kuala Lumpur}
  \country{Malaysia}
}


\renewcommand{\shortauthors}{Cai et al.}
\begin{abstract}

Recent advancements in information retrieval have highlighted the potential of integrating visual and textual information, yet effective reranking for image-text documents remains challenging due to the modality gap and scarcity of aligned datasets. Meanwhile, existing approaches often rely on large models (7B–32B parameters) with reasoning-based distillation, incurring unnecessary computational overhead while primarily focusing on textual modalities. In this paper, we propose \textbf{Rank-Nexus}, a multimodal image-text document reranker that performs listwise qualitative reranking on retrieved lists incorporating both images and texts. To bridge the modality gap, we introduce a progressive cross-modal training strategy. We first train modalities separately: leveraging abundant text reranking data, we distill knowledge into the text branch. For images, where data is scarce, we construct distilled pairs from multimodal large language model (MLLM) captions on image retrieval benchmarks. Subsequently, we distilled a joint image-text reranking dataset. Rank-Nexus achieves outstanding performance on text reranking benchmarks (TREC, BEIR) and the challenging image reranking benchmark (INQUIRE, MMDocIR), using only a lightweight 2B pretrained visual-language model. This efficient design ensures strong generalization across diverse multimodal scenarios without excessive parameters or reasoning overhead.
\end{abstract}



\begin{teaserfigure}
  \includegraphics[width=1.0 \textwidth]{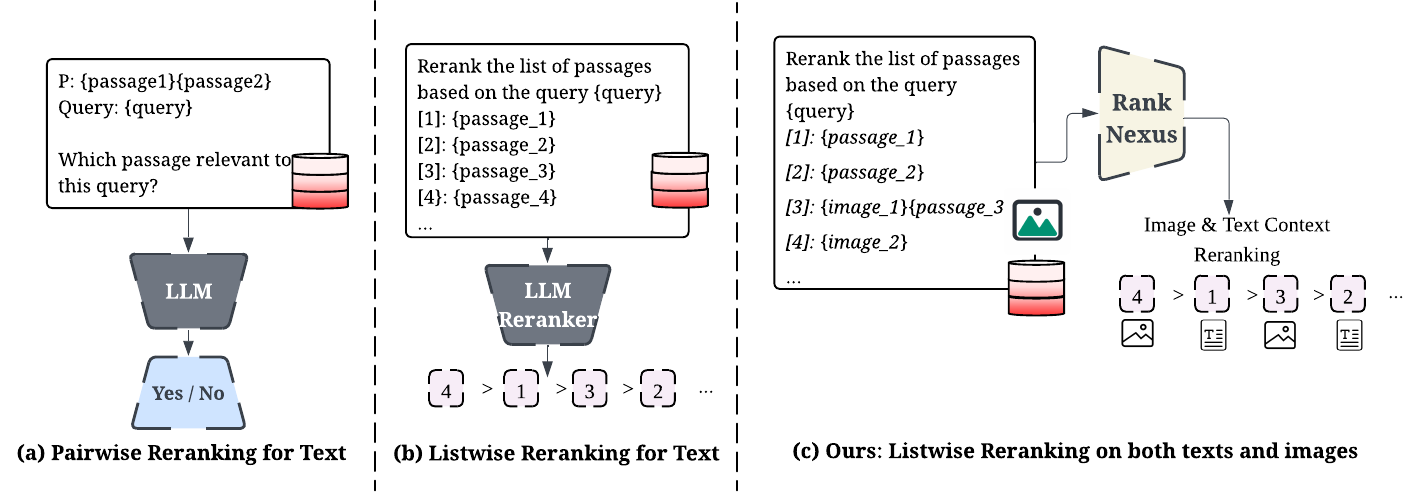}
  \caption{Comparison of three reranking paradigms. (a) Pairwise reranking~\cite{nogueira2020document, qin2023large} evaluates each passage independently against the query using an LLM to determine relevance with binary yes/no decisions. (b) Listwise reranking~\cite{ma2023zero, pradeep2023rankvicuna} processes all candidate passages simultaneously, and directly outputs them in sorted ranking. (c) Our proposed Rank-Nexus approach extends listwise reranking to multimodal scenarios, jointly processing both textual passages and images to generate comprehensive rankings that leverage visual and textual information.}
  \Description{Comparison of three reranking paradigms for information retrieval.}
  \label{fig:teaser}
\end{teaserfigure}

\maketitle


\section{Introduction}

\textit{"A picture is worth a thousand words,"} -- advertising executive Frederick R. Barnard.

Large Language Models (LLMs)~\cite{brown2020language, touvron2023llama2, Touvron2023Llama2O} have been proven effective document rerankers from the advantages of its strong context understanding and pre-training prior knowledge. Using LLMs to rank document with specific instructions is categorized into two downstream tasks: pairwise reranking~\cite{nogueira2020document, qin2023large} and listwise reranking~\cite{ma2023zero, pradeep2023rankvicuna}. Both settings in tandem leverages LLMs output and format them with patterns, which pairwise reranking focuses on single question relativity identification, while listwise reranking involves LLMs with long contexts to select the ranking for each document. However, most data today are documents mixed with tables, illustration and captions. While these approaches have advanced textual retrieval, their extension to multimodal documents that have both text and visual elements, remains fundamentally limited, as illustrated in Fig. \ref{fig:teaser}.

There are several challenges of training a Vision-Language Model (VLM) reranker at the moment; First, the gap needs to be bridged between images and texts to understand each priority relevant to the query, as how the VLM model process multimodal data in the same embedding space to achieve reranking~\cite{team2023gemini}. More radicially, since most popular methods of reranking models are based on text~\cite{lin2021pretrained}, distilling from MS-MARCO~\cite{bajaj2016msmarco}, ignoring visual content like illustration and tables, the datasets for both training and evaluating multimodal information retrieval are scarce, especially on listwise reranking, which focuses more on the overall quality of ranking results. Although it is still possible to re-rank documents without pre-training, the large burdening computational cost and inference time from listwise reranking of text results in inefficiency, let alone image involved reranking, nearly impossible to be deployed in production. Therefore, we found and aim to solve one question: \textbf{\textit{Where can we seek the trade-off on reranking documents, each contains a vast amount of pictures and words?}}

To address data issues, we propose a complete further data distillation pipeline along with efficient coreset selection on reranking training data, as well as integrating image-text multimodal documents from one of popular document retrieval dataset, MMDocIR~\cite{dong2025mmdocir}. We demonstrate that only the core part of reranking training data are significant to be involved, although the existing dataset RankGPT~\cite{ma2023zero} have been extending to 10k scale. We also observe the diminishing return of dataset up-scaling and adopted greedy diversity-based searching strategy for fine-graining qualitative data to reduce unnecessary computation, which is agnostic to the modality--used in both text and images. Therefore, only 7.5\% text information retrieval and image information retrieval (1000 pairwise samples + 2100 listwise samples) are used during training.

\begin{figure}
    \centering
    \includegraphics[width=1.0\linewidth]{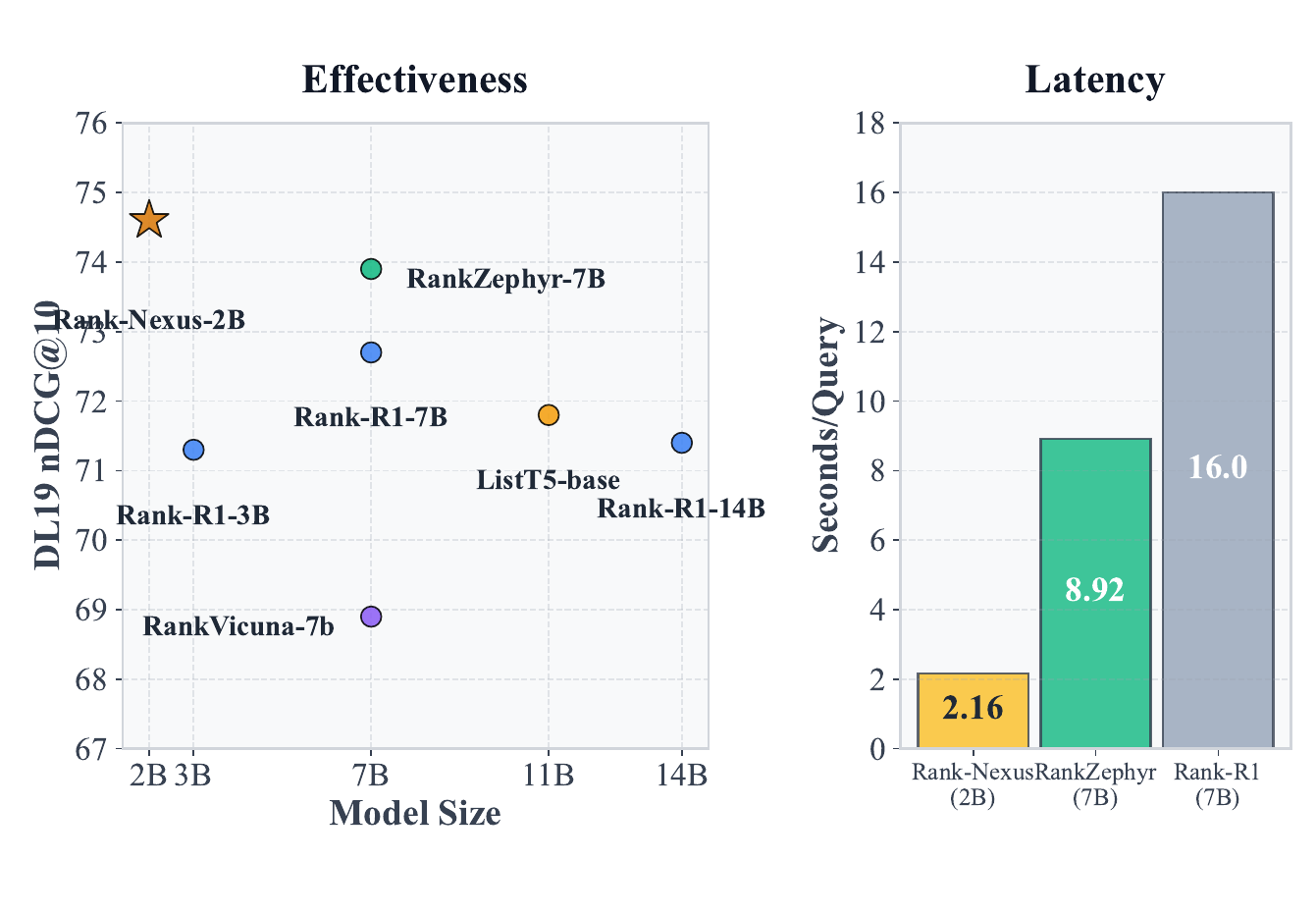}
    \caption{The effectivenss (left) and latency (right) of Rank-Nexus and other baseline method on document listwise reranking inference. We experimented and illustrated DL-19 benchmark for evaluation and one of distilled MSMARCO dataset as inference time example. Rank-R1 takes most latency among 3 models, strongly dependent on reasoning time. While Rank-Nexus are the fastest as it possesses smallest parameter size and outputs without ad-hoc thinking.}
    \label{fig:fig2}
\end{figure}

Building upon this foundation of high-quality multimodal ranking data, we introduce a progressive training strategy to train our VLM reranker model, \textbf{Rank-Nexus}, which systematically transitions the model from text-only to multimodal listwise reranking. This curriculum progresses through three stages: the model first learns fundamental ranking on pairwise text data, then acquires basic visual-text alignment through pairwise image ranking, and finally masters global document ordering via listwise image ranking. Crucially, our approach departs from currently trending reasoning-based rerankers, Rank-R1~\cite{zhuang2025rankr1} built upon 7B+ LLMs, whose extensive inference overhead often negates their marginal relevance gains. Instead, our compact 2B multimodal reranker operates through direct forward computation without expensive reasoning chains~\cite{kojima2022large}, achieving superior inference efficiency—completing MSMARCO passage reranking in just 2 seconds compared to 8.92 seconds (RankZephyr~\cite{pradeep2023rankzephyr}) and 16 seconds (RankR1)—while simultaneously attaining higher ranking accuracy, as quantitatively demonstrated in Fig. \ref{fig:fig2}. This synergistic combination of progressive multimodal training, lightweight architecture, and efficient inference enables state-of-the-art reranking performance with dramatically reduced computational requirements across both training and deployment.

We conduct the extensive experiment to verify the effectiveness of Rank-Nexus ranking on both text ranking benchmark (DL19~\cite{craswell2020overview}, DL20~\cite{craswell2021overview}) and image ranking benchmark (INQUIRE~\cite{vendrow2024inquire}, MMDocIR~\cite{dong2025mmdocir}). Our 2B model outperforms the majority of open-source distilled instruction models. Our contribution can be summarized below:
\begin{itemize}
    \item We introduce Rank-Nexus, the lightweight model for end-to-end listwise reranking of mixed image-text documents, directly optimizing global ordering in a unified framework.
    \item We develop a data curation and progressive training strategy that achieves strong performance using minimal high-quality data, reducing training costs.
    \item Our model enables fast inference through direct computation, significantly outperforming reasoning-based rerankers in speed while maintaining competitive accuracy on benchmarks.
\end{itemize}

\section{Related Work}

\subsection{Document Reranking}
Document reranking has evolved from traditional learning-to-rank methods~\cite{liu2009learning} to modern neural approaches leveraging pre-trained language models~\cite{lin2021pretrained}. Early neural rerankers employed direct scoring based on points~\cite{nogueira2020document, zhuang2021tilde} that evaluates each document independently, while pairwise methods~\cite{pradeep2021expando, qin2023large} compare document pairs for relative relevance. Recent advances focus on listwise reranking~\cite{ma2023zero, pradeep2023rankvicuna}, which processes entire candidate lists simultaneously to optimize global ranking quality. The emergence of Large Language Models has further revolutionized this field, with zero-shot approaches~\cite{sun2023chatgpt, ma2023zero} demonstrating strong performance through instruction-following capabilities. Subsequent work has distilled these capabilities into smaller models: RankVicuna~\cite{pradeep2023rankvicuna} and RankZephyr~\cite{pradeep2023rankzephyr} achieve competitive performance with 7B parameters, while reasoning-based rerankers like Rank-R1~\cite{zhuang2025rankr1} incorporate explicit reasoning chains at the cost of substantial computational overhead. However, all existing reranking methods focus exclusively on textual content, ignoring rich visual information in modern documents such as figures, tables, and layout structures.

\subsection{Multimodal Retrieval}
Vision-language models have transformed multimodal retrieval through large-scale contrastive learning between images and text. CLIP~\cite{radford2021learning} and ALIGN~\cite{jia2021scaling} pioneered this paradigm by training dual encoders on massive image-text pairs, creating shared embedding spaces for zero-shot retrieval. Subsequent work refined these approaches: SigLIP~\cite{zhai2023sigmoid} introduced sigmoid loss for improved training stability, SigLIP 2~\cite{siglip2_2025} enhanced semantic understanding and localization, while ALBEF~\cite{li2021align} proposed aligning unimodal representations before fusion with momentum distillation. BLIP~\cite{li2022blip} unified understanding and generation through multi-task learning with caption filtering, and BLIP-2~\cite{li2023blip2} efficiently leveraged frozen encoders via lightweight Q-Former, achieving state-of-the-art performance with 54× fewer parameters. For document-specific scenarios, INQUIRE~\cite{vendrow2024inquire} and MMDocIR~\cite{dong2025mmdocir} benchmarks revealed significant challenges in retrieving visually rich documents, while ColPali~\cite{faysse2024colpali} demonstrated that directly embedding page images with late interaction outperforms complex OCR pipelines. Despite these advances in first-stage multimodal retrieval, the problem of multimodal reranking remains unexplored, with existing VLMs like Qwen2-VL~\cite{wang2024qwen2vl} and Gemini~\cite{team2023gemini} designed for understanding rather than listwise reranking. Our work addresses this gap by introducing lightweight VLM-based listwise reranker for mixed image-text documents.

\section{Methodology}

\label{sec:method}

\begin{figure*}
    \centering
    \includegraphics[width=0.97\linewidth]{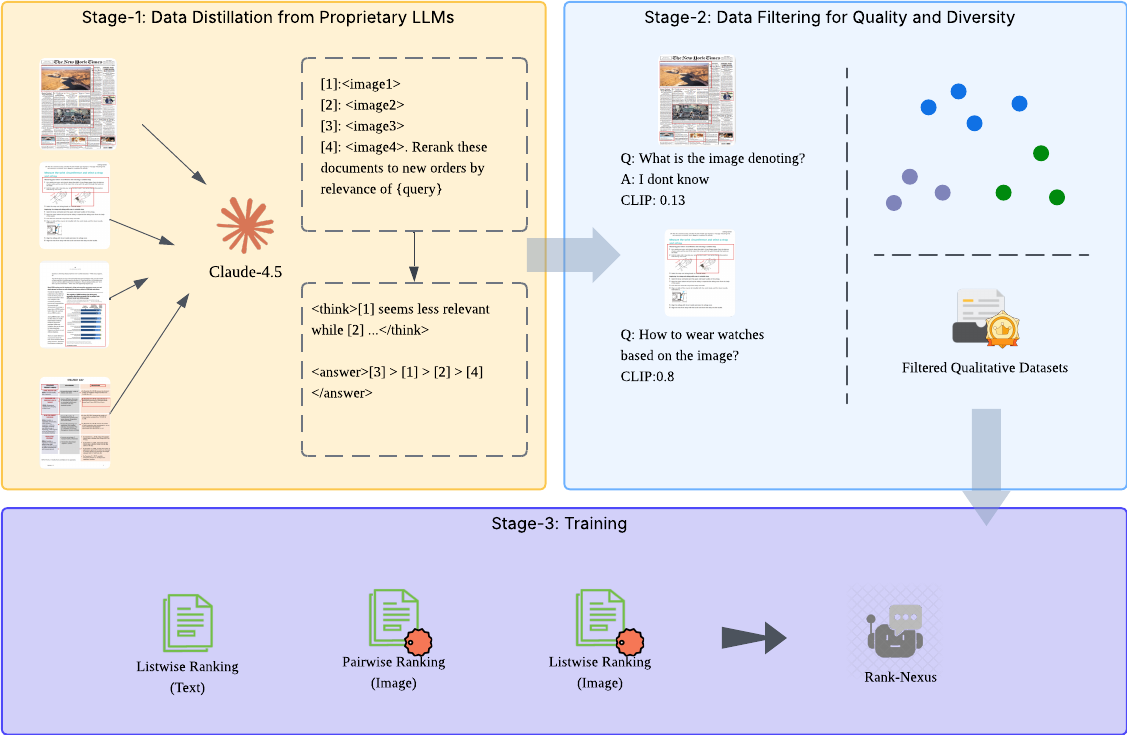}
    \caption{Overview of Rank-Nexus training pipeline. \textbf{Stage-1}: We distill listwise rankings from Claude-4.5 on diverse multimodal documents, generating structured output with relevance ordering and explanations. \textbf{Stage-2}: Quality filtering removes low-confidence samples (CLIP score threshold) while diversity-based coreset selection ensures coverage across semantic clusters, producing high-quality training data. \textbf{Stage-3}: Progressive curriculum trains the model sequentially on text listwise ranking, image pairwise ranking, and image listwise ranking to build comprehensive multimodal reranking capabilities.}
    \label{fig:placeholder}
\end{figure*}
\subsection{Preliminary}

\subsubsection{Problem Formulation}

Given a query $q$ and a candidate document set $\mathcal{D} = \{d_1, d_2, \ldots, d_n\}$ retrieved by a first-stage retriever, where each document $d_i$ may contain text, images, or both, our goal is to produce a permutation $\pi: [n] \rightarrow [n]$ that reorders $\mathcal{D}$ by relevance to $q$. We parameterize this ranking function with a vision-language model $\theta$ that processes the query and all candidate documents to output the optimal ordering.

\subsubsection{Pairwise Reranking}
In the pairwise approach~\cite{qin2024largelanguagemodelseffective}, the model learns to compare pairs of documents given a query. For each query $q$ and a pair of documents $(d_i, d_j)$, the model predicts which document is more relevant:
\begin{equation}
\text{rank}(d_i) = \sum_{j \neq i} \mathbb{1}[P(d_i \succ d_j | q; \theta) > 0.5]
\end{equation}
where $\mathbb{1}[\cdot]$ is the indicator function. The final ranking of pairwise results are from simply sorting each document by its accumulated wins over other documents.

\subsubsection{Listwise Ranking}

Unlike pairwise approaches, listwise ranking~\cite{learningtorank} processes all candidate documents simultaneously. Given query $q$ and document list $\mathcal{D}$, the model outputs relevance scores for all documents:
\begin{equation}
\{s_\theta(q, d_1), s_\theta(q, d_2), \ldots, s_\theta(q, d_n)\} = f_\theta(q, \mathcal{D})
\end{equation}

Following the Plackett-Luce model~\cite{xia2019plackettlucemodellearningtoranktask}, the probability of observing a ground-truth permutation $\pi^*$ is:
\begin{equation}
P(\pi^* | q, \mathcal{D}; \theta) = \prod_{i=1}^{n} \frac{\exp(s_\theta(q, d_{\pi^*(i)}))}{\sum_{j=i}^{n} \exp(s_\theta(q, d_{\pi^*(j)}))}
\end{equation}

The corresponding listwise ranking loss with temperature $\tau$ is:
\begin{equation}
\mathcal{L}_{\text{listwise}}^\tau = -\sum_{i=1}^{n} \log \frac{\exp(s_\theta(q, d_{\pi^*(i)}) / \tau)}{\sum_{j=i}^{n} \exp(s_\theta(q, d_{\pi^*(j)}) / \tau)}
\label{eq:listwise_loss}
\end{equation}

\subsubsection{Vision-Language Embeddings}

We utilize CLIP~\cite{radford2021learning} to obtain aligned visual-textual representations. For an image $I$ and text $t$, CLIP provides:
\begin{equation}
\mathbf{e}_I = \text{Enc}_{\text{img}}(I), \quad \mathbf{e}_t = \text{Enc}_{\text{text}}(t)
\end{equation}
where both encoders project inputs into a shared $d$-dimensional embedding space. Similarity is computed via cosine similarity:
\begin{equation}
\text{sim}(I, t) = \frac{\mathbf{e}_I \cdot \mathbf{e}_t}{\|\mathbf{e}_I\|_2 \|\mathbf{e}_t\|_2}
\end{equation}

For efficiency in large-scale retrieval, we also use Euclidean distance:
\begin{equation}
d(I, t) = \|\mathbf{e}_I - \mathbf{e}_t\|_2
\label{eq:euclidean}
\end{equation}

\subsection{Overview}

We propose Rank-Nexus, a 2B-parameter multimodal listwise reranker that processes both textual and visual documents. Our approach employs a progressive cross-modal training strategy: we first train on text and image modalities separately, then perform joint multimodal fine-tuning. Unlike prior work that trains on 100K+ samples, we demonstrate that strategic data selection enables superior performance with only 7.5\% of text training data and 3\% of image training data.

\subsection{Text Reranking with Knowledge Distillation}

We begin by training the text branch through knowledge distillation from proprietary LLMs. For each query $q$ and passage list $\mathcal{P} = \{p_1, p_2, \ldots, p_n\}$, we query teacher models (GPT-4, Claude-3.5-Sonnet) to generate optimal rankings $\pi^*$. Our model learns to minimize the listwise ranking loss:
\begin{equation}
\mathcal{L}_{\text{text}} = -\sum_{i=1}^{n} \log \frac{\exp(s_\theta(q, p_{\pi^*(i)}) / \tau)}{\sum_{j=i}^{n} \exp(s_\theta(q, p_{\pi^*(j)}) / \tau)}
\end{equation}
where $\tau = 0.1$ is the temperature parameter.

\subsubsection{Data Quality over Quantity}

We conduct ablation studies with training sets of varying sizes on TREC DL19 and DL20. Results show performance peaks at 4,000 samples, with larger datasets degrading performance. This indicates that beyond a certain point, additional data introduces noise and redundancy rather than useful signal. We employ confidence-based filtering to select the top 4,000 highest-quality samples from teacher model rankings, representing only 7.5\% of the 100K samples used by baseline methods. Further case studies and ablation experiments can be referred to Chapter \ref{casestudy}.

\subsection{Image Reranking}
After gaining reranking ability that migrated from text document retrieval, we continue the initialized model weights to train image ranking.
Training an image reranking model faces challenges due to the scarcity of listwise image ranking datasets. We construct training data from MMDocIR~\cite{dong2025mmdocir}, a multimodal document retrieval benchmark containing 1685 QA pairs from 313 documents with 173,843 questions with bootstrapped labels.

\textbf{Motivation.} Rather than using all 73K samples, we argue that indiscriminate training on large datasets leads to overfitting, computational waste, and semantic homogeneity. Instead, we propose a dataset selection framework that prioritizes semantic diversity, removes redundancy, and ensures high query-image relevance. This allows us to distill a compact set of informative samples that better represent the underlying distribution.

We apply three filtering and selection strategies:

\paragraph{Quality-Aware Filtering.}
We first filter out low-quality query-image pairs using CLIP-based relevance scoring. For each pair $(q, I)$, we compute the cosine similarity between their embeddings and retain only pairs with strong similarity to make sure remove weakly-related pairs that would introduce noisy supervision signals.

\paragraph{Maximum Diversity Selection.}
After quality filtering, we select a diverse subset that maximizes coverage of the visual semantic space. Our motivation is to maximize inter-document diversity to reduce redundancy in between data. Inspired from DivPrune~\cite{divprune}, a diversity-based visual token pruning method for large multimodal models, selecting the most diverse subset is identified an instance of the Maximum Marginal Diversity Problem (MMDP), which is NP-hard. Therefore, we adjusted token-level submodular algorithm into document level, which we employ a greedy approximation that iteratively selects documents most dissimilar to the current selection.

The algorithm works as follows: we start with an initial document and iteratively add the document that has the lowest average similarity to all currently selected documents. This ensures each newly added document provides maximum marginal diversity. The process continues until we reach the target subset size $k$. Algorithm~\ref{alg:diversity} presents the complete procedure.

\begin{algorithm}[t]
\caption{Greedy Maximum Diversity Selection}
\label{alg:diversity}

\begin{algorithmic}[1]
\REQUIRE Filtered candidate set $\mathcal{V} = \{v_1, \ldots, v_N\}$ with CLIP embeddings $\{\mathbf{e}_1, \ldots, \mathbf{e}_N\}$, target size $k$
\ENSURE Diverse subset $\mathcal{S}$ with $|\mathcal{S}| = k$
\STATE Initialize $\mathcal{S} \leftarrow \{v_1\}$
\FOR{$t = 2$ \TO $k$}
    \FOR{each candidate $v_j \in \mathcal{V} \setminus \mathcal{S}$}
        \STATE Compute $\text{avg\_sim}(v_j, \mathcal{S}) = \frac{1}{|\mathcal{S}|} \sum_{v_i \in \mathcal{S}} \frac{\mathbf{e}_i \cdot \mathbf{e}_j}{\|\mathbf{e}_i\|_2 \|\mathbf{e}_j\|_2}$
    \ENDFOR
    \STATE $v^* \leftarrow \arg\min_{v_j \in \mathcal{V} \setminus \mathcal{S}} \text{avg\_sim}(v_j, \mathcal{S})$
    \STATE $\mathcal{S} \leftarrow \mathcal{S} \cup \{v^*\}$
\ENDFOR
\RETURN $\mathcal{S}$
\end{algorithmic}
\end{algorithm}

\paragraph{Dataset Distillation with Efficient Processing.}
To construct the final training set, We first load the MMDocIR Parquet dataset and precompute CLIP embeddings for all images and queries. We leverage DuckDB for scalable processing since the parquet files are column-based serially stored in the disk. For each query, we retrieve the top-20 most relevant images based on Euclidean distance in the embedding space. We prompt Claude-4.5-Haiku to generate rerankings for each query and its top-20 candidates, producing distilled training labels. This pipeline processes the entire 73K dataset and extracts 2,100 high-quality training samples (3\% of original data) in approximately 2 hours on a single GPU.

\subsection{Joint Multimodal Fine-Tuning}

After training separate text and image branches, we perform joint fine-tuning on mixed image-text document lists. We initialize from the image-trained checkpoint, which already incorporates text knowledge, and continue training on mixed-modality data. The multimodal ranking objective is:
\begin{equation}
\mathcal{L}_{\text{mm}} = -\sum_{i=1}^{m} \log \frac{\exp(s_\theta(q, d_{\pi^*(i)}) / \tau)}{\sum_{j=i}^{m} \exp(s_\theta(q, d_{\pi^*(j)}) / \tau)}
\end{equation}
where each document $d_i$ in list $\mathcal{M}$ can be text, image, or hybrid. We generate pseudo-labels for mixed lists using the same LLM distillation approach.

\section{How Does Data Affect Reranking Results?}
\label{casestudy}

Training data plays a critical role in determining reranking model performance, but the relationship between data scale and effectiveness is not straightforward. In this section, we investigate two key aspects: (1) how text data quantity impacts ranking quality, revealing a diminishing return phenomenon, and (2) how progressive multimodal training incrementally builds image understanding capabilities on top of text reranking foundations.

\subsection{Diminishing Returns on Text Data Scaling}

We systematically examine the relationship between text training data scale and reranking performance by fine-tuning Rank-Nexus on different subsets of the MS MARCO passage ranking dataset. Figure~\ref{fig:placeholder} illustrates model effectiveness on TREC DL19 and DL20 benchmarks across four scales: 1k, 4k, 8k, and 10k training samples.

The results reveal a clear diminishing return pattern. Performance improves sharply from 1k to 4k samples, with DL19 nDCG@10 increasing from 70.8 to 74.9—a gain of 4.1 points. However, further scaling to 8k and 10k samples not only fails to improve but actually degrades performance, declining to 73.2 and 73.1 respectively. This trend appears consistently on DL20, where performance peaks at 70.0 (4k samples) before dropping to 69.2 (8k) and 69.4 (10k).

Table~\ref{tab:data_quantity_text} quantifies these observations. The 4k configuration represents the optimal scale, marked as "Optimal" in Figure~\ref{fig:placeholder}. Beyond this point, additional data introduces diminishing returns due to three factors: (1) redundancy—new samples fail to provide novel ranking patterns, (2) noise accumulation—larger datasets inevitably contain lower-quality annotations that dilute the learning signal, and (3) overfitting risk—extended training on larger datasets reduces generalization to unseen test distributions.

\begin{table}[h]
\centering
\setlength{\abovecaptionskip}{0.05in} 
\setlength{\belowcaptionskip}{0.05in}
\caption{Impact of text training data quantity on passage reranking effectiveness. Performance is measured by nDCG@10 on TREC DL19 and DL20 test sets.}
\label{tab:data_quantity_text}
\renewcommand{\arraystretch}{1.1}
\setlength{\tabcolsep}{6pt}
\small
\begin{tabular}{lcccr}
\toprule
\textbf{Data Scale} & \textbf{Samples} & \textbf{DL19} $\uparrow$ & \textbf{DL20} $\uparrow$ & \textbf{Avg.} $\uparrow$ \\
\midrule
Small & 1k & 70.8 & 68.1 & 69.5 \\
\textbf{Optimal} & \textbf{4k} & \textbf{74.9} & \textbf{70.0} & \textbf{72.5} \\
Large & 8k & 73.2 & 69.2 & 71.2 \\
Full & 10k & 73.1 & 69.4 & 71.3 \\
\bottomrule
\end{tabular}
\end{table}

Our results demonstrate that strategic data selection at moderate scales (4k samples) outperforms naive scaling to larger datasets. For resource scenarios, we suggest prioritizing curating high-quality, diverse training samples rather than maximizing dataset size. 

\subsection{Progressive Multimodal Training Builds Image Understanding}

While text reranking provides a strong foundation, extending these capabilities to image and multimodal document ranking requires careful curriculum design. We analyze how our three-stage progressive training strategy incrementally improves performance on image-centric benchmarks: INQUIRE (text-to-image retrieval) and MMDocIR (multimodal document retrieval).

Table~\ref{tab:progressive_training_stages} presents results across four training stages: (1) no training (baseline retriever only), (2) after text reranking training, (3) after adding Pairwise image reranking, and (4) after adding listwise image reranking (our full model). The progression reveals how each training stage contributes to the final multimodal ranking capability.

\begin{table}[h]
\centering
\setlength{\abovecaptionskip}{0.05in} 
\setlength{\belowcaptionskip}{0.05in}
\caption{Progressive training stages and their impact on image-centric reranking tasks. We show how text training, Pairwise image training, and listwise image training incrementally build multimodal ranking capabilities.}
\label{tab:progressive_training_stages}
\renewcommand{\arraystretch}{1.1}
\setlength{\tabcolsep}{5pt}
\small
\begin{tabular}{lcc}
\toprule
\textbf{Training Stage} & \makecell{\textbf{INQUIRE} \\ \textbf{(nDCG@50)} $\uparrow$} & \makecell{\textbf{MMDocIR} \\ \textbf{(Recall@1)} $\uparrow$} \\
\midrule
Baseline (No training) & 60.1 & 68.8 \\
+ Text reranking training & 62.9 & 69.1 \\
+ Pairwise image reranking & 63.2 & 69.3 \\
\midrule
\textbf{+ Listwise image reranking (Full)} & \textbf{69.0} & \textbf{70.2} \\
\bottomrule
\end{tabular}
\end{table}

\textbf{Stage 1: Text reranking foundation.} Surprisingly, text reranking training alone improves image ranking performance, with INQUIRE increasing from 60.1 to 62.9 (+2.8) and MMDocIR from 68.8 to 69.1 (+0.3). This unexpected benefit arises because text reranking teaches the model fundamental ranking principles—understanding relevance signals, comparing candidates, and producing coherent orderings. These learned capabilities transfer to the visual domain, even though no image data is involved at this stage. Additionally, many image queries in these benchmarks contain textual metadata or captions that benefit from improved text understanding.

\textbf{Stage 2: Pairwise image understanding.} Adding Pairwise image reranking training provides modest gains (INQUIRE: 62.9 → 63.2, MMDocIR: 69.1 → 69.3). This stage trains the model to assess individual image-query relevance independently. While this improves basic visual understanding, Pairwise training lacks the global context needed to distinguish between multiple similarly relevant images, limiting its effectiveness for complex ranking scenarios.

\textbf{Stage 3: Listwise image reranking.} The final stage, listwise image reranking training, delivers substantial improvements (INQUIRE: 63.2 → 69.0, MMDocIR: 69.3 → 70.2). The 5.8-point gain on INQUIRE is particularly striking, representing a 9.2\% relative improvement over the Pairwise baseline. Listwise training enables the model to perform global comparisons across all candidates simultaneously, learning to distinguish fine-grained visual differences and optimize the overall ranking quality rather than isolated relevance scores.

The cumulative gains across stages are remarkable: from baseline to full model, INQUIRE improves by 8.9 points (60.1 → 69.0, +14.8\%) and MMDocIR by 1.4 points (68.8 → 70.2, +2.0\%). This progressive training strategy is essential—our ablations in Section 5 show that directly training on joint image-text data without this curriculum yields significantly worse results. The curriculum allows the model to first establish strong textual ranking capabilities, then layer on visual understanding through Pairwise training, and finally master global multimodal ranking through listwise training.

These findings demonstrate that multimodal reranking capabilities emerge progressively rather than spontaneously. Each training stage builds upon previous foundations, with the final listwise image training being the critical component that unlocks strong image ranking performance. This validates our progressive training design and provides a practical roadmap for training efficient multimodal rerankers.
\begin{figure}
    \centering
    \includegraphics[width=1\linewidth]{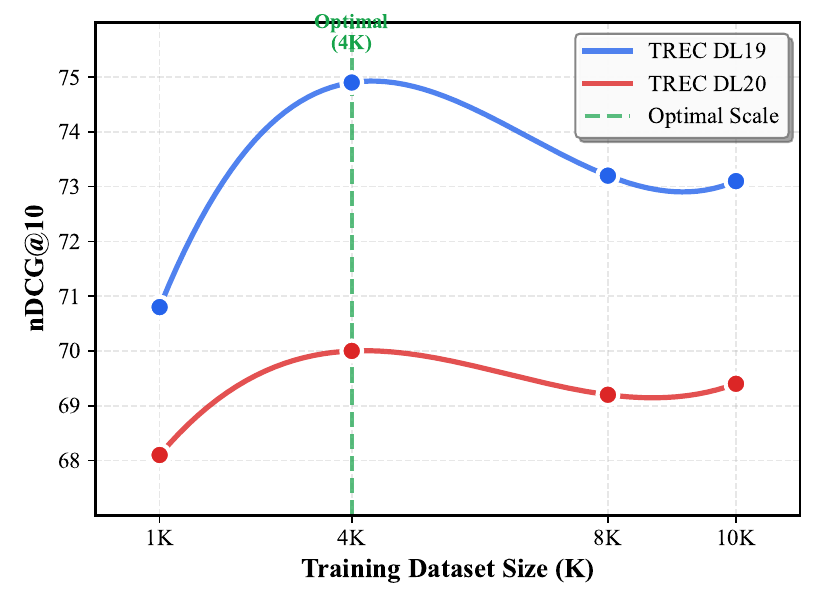}
    \caption{Diminishing return on data scaling. We have experimented on 1k, 4k, 8k and 10k of MSMARCO training dataset for fine-tuning in Rank-Nexus-2B-Qwen.}
    \label{fig:placeholder}
\end{figure}

\vspace{1em}

\begin{table*}[t]
\centering
\setlength{\abovecaptionskip}{0.05in} 
\caption{Results (NDCG@10) on 7 BEIR datasets. All models rerank BM25-retrieved top-100 passages. \textbf{Bold} and \underline{underline} indicate the best and the second best results, respectively. \colorbox{gray!20}{Gray background} denotes our proposed methods.}
\label{tab:beir_results}
\renewcommand{\arraystretch}{1.2}
\setlength{\tabcolsep}{5pt}
\footnotesize
\resizebox{\textwidth}{!}{%
\begin{tabular}{l|cccccccc}
\toprule
\textbf{Models} & \textbf{Covid} $\uparrow$ & \textbf{DBPedia} $\uparrow$ & \textbf{SciFact} $\uparrow$ & \textbf{NFCorpus} $\uparrow$ & \textbf{Signal} $\uparrow$ & \textbf{Robust04} $\uparrow$ & \textbf{News} $\uparrow$ & \textbf{Avg.} $\uparrow$ \\
\midrule
BM25 & 59.47 & 31.80 & 67.89 & 33.75 & 33.04 & 40.70 & 39.52 & 43.74 \\
\midrule
\multicolumn{9}{l}{\textit{Non-reasoning reranker}} \\
RankT5 (3B) & 80.19 & 44.85 & 74.60 & 37.40 & 31.73 & 51.45 & 49.11 & 52.76 \\
RankZephyr (7B) & 84.00 & 44.60 & \underline{74.90} & 36.70 & 31.80 & 54.30 & \underline{52.60} & 54.13 \\
ListT5-base & 78.30 & 43.70 & 74.10 & 35.60 & 33.50 & 52.10 & 48.50 & 52.26 \\
\midrule
\multicolumn{9}{l}{\textit{Reasoning reranker}} \\
Rank-R1 (7B) & 83.71 & 42.27 & 72.16 & \textbf{38.94} & 33.08 & 54.46 & 50.60 & 53.60 \\
Rank-R1 (14B) & \underline{84.63} & 44.05 & \textbf{75.96} & \underline{38.58} & 32.95 & \underline{56.91} & 49.20 & \underline{54.61} \\
Rank1 (7B) & 79.04 & 35.79 & 73.32 & 37.52 & 25.41 & \textbf{57.11} & 47.67 & 50.84 \\
RankVicuna-7b & 80.50 & 44.40 & 70.80 & 33.20 & \textbf{34.20} & 48.90 & 46.90 & 51.27 \\
RankGPT(GPT-3.5) & 76.70 & 44.50 & 70.40 & 35.60 & 32.10 & 50.60 & 48.90 & 51.26 \\
\midrule
\rowcolor{gray!15}
Rank-Nexus-2B (QwenVL) & 80.22 & \underline{45.10} & 73.84 & 36.75 & 33.41 & 53.47 & \underline{52.60} & 50.27 \\
\rowcolor{gray!15}
Rank-Nexus-2B (InternVL) & \textbf{85.99} & \textbf{45.70} & 72.21 & 37.60 & \underline{34.17} & 55.72 & \textbf{53.11} & \textbf{54.92} \\
\bottomrule
\end{tabular}%
}
\end{table*}

\vspace{1em}

\vspace{1em}

\begin{table*}[t]
\centering
\setlength{\abovecaptionskip}{0in} 
\setlength{\belowcaptionskip}{0in}
\caption{Performance on MMDocIR. "Retriever-only" indicates using retrieval results directly as final output, while "Reranking from ColQwen" means reranking the top candidates retrieved by ColQwen. Methods with "cot" suffix generate chain-of-thought reasoning during reranking. Bold indicates best performance}
\label{tab:mmdocir_results}
\renewcommand{\arraystretch}{1.2}
\setlength{\tabcolsep}{4pt}
\footnotesize
\resizebox{\textwidth}{!}{%
\begin{tabular}{ll|ccc|ccc}
\toprule
\multirow{2}{*}{\textbf{Algorithm}} & \multirow{2}{*}{\textbf{Method}} & \multicolumn{3}{c|}{\textbf{Macro}} & \multicolumn{3}{c}{\textbf{Micro}} \\
\cmidrule(lr){3-5} \cmidrule(lr){6-8}
& & \textbf{Recall@1} & \textbf{Recall@3} & \textbf{Recall@5} & \textbf{Recall@1} & \textbf{Recall@3} & \textbf{Recall@5} \\
\midrule
\multirow{5}{*}{Retriever-only} 
& CLIP~\cite{radford2021learningtransferablevisualmodels} & 0.3334 & 0.5428 & 0.6452 & 0.3227 & 0.5337 & 0.6350 \\
& E5-V~\cite{jiang2024e5vuniversalembeddingsmultimodal} & 0.4201 & 0.6470 & 0.7389 & 0.4249 & 0.6578 & 0.7519 \\
& DSE~\cite{ma2024unifyingmultimodalretrievaldocument} & 0.5109 & 0.7194 & 0.7925 & 0.5023 & 0.7234 & 0.7971 \\
& GME~\cite{zhang2025gmeimprovinguniversalmultimodal} & 0.5400 & 0.7603 & 0.8308 & 0.5421 & 0.7603 & 0.8377 \\
& ColQwen~\cite{faysse2024colpali} & 0.6481 & 0.8331 & 0.8766 & 0.6354 & 0.8213 & 0.8667 \\
\midrule
\multirow{10}{*}{Reranking from ColQwen}
& RagVL~\cite{chen2024mllmstrongrerankeradvancing} & 0.3814 & 0.6462 & 0.7667 & 0.3411 & 0.6206 & 0.7497 \\
& Qwen2.5-VL-7B~\cite{wang2024qwen2vl} & 0.6230 & 0.8243 & 0.8675 & 0.6012 & 0.8069 & 0.8555 \\
& Qwen2.5-VL-7B-cot & 0.6479 & 0.8179 & 0.8670 & 0.6336 & 0.8103 & 0.8604 \\
& Gemma3-12B~\cite{gemmateam2025gemma3technicalreport} & 0.4743 & 0.6675 & 0.7278 & 0.4456 & 0.6353 & 0.7025 \\
& Gemma3-12B-cot & 0.5403 & 0.7729 & 0.8292 & 0.5166 & 0.7536 & 0.8153 \\
& Qwen2.5-VL-7B-sft & 0.6673 & 0.8475 & 0.8828 & 0.6498 & 0.8366 & 0.8746 \\
& MM-R5~\cite{xu2025mmr5multimodalreasoningenhancedreranker} & \textbf{0.6951} & \textbf{0.8520} & \textbf{0.8842} & \textbf{0.6759} & \textbf{0.8401} & \textbf{0.8755} \\

& Rank-Nexus (ours) & \textbf{0.7010} & \textbf{0.8696} & \textbf{0.8753} & \textbf{0.6710} & \textbf{0.8497} & \textbf{0.8700} \\
\bottomrule
\end{tabular}%
}
\vspace{0em}
\end{table*}

\section{Experiments}

\subsection{Datasets}

We evaluate Rank-Nexus on both text-only and multimodal benchmarks to demonstrate its effectiveness across diverse retrieval scenarios.

\textbf{MS MARCO Passage Ranking}~\cite{bajaj2016msmarco} is a large-scale information retrieval dataset containing 8.8 million passages and over 500,000 training queries derived from real Bing search logs. The dataset provides sparse relevance judgments (typically 1-2 relevant passages per query) and is widely used for training neural ranking models. We use the standard passage reranking setting where models rerank the top-1000 candidates retrieved by BM25.

\textbf{TREC DL19 and DL20}~\cite{craswell2020overview, craswell2021overview} are test collections from the TREC Deep Learning Track 2019 and 2020, designed to evaluate ad hoc passage ranking in the large data regime. DL19 comprises 43 test queries while DL20 contains 54 queries, both sharing the MS MARCO passage corpus of 8.8 million passages. Unlike MS MARCO's sparse judgments, these datasets feature comprehensive NIST human annotations with multi-level relevance grades. The official evaluation metric is NDCG@10, which we report in our experiments.

\begin{table}[h]
\centering
\setlength{\abovecaptionskip}{0in} 
\setlength{\belowcaptionskip}{0in}
\caption{TREC DL19 and DL20 nDCG@10 results with dataset scale and average scores. Here, zeroshot denotes model inference without any fine-tuning. LLM-distil and SFT denotes it is distilled from large language model outputs. And RL denotes the model is fine-tuned through reinforcement training.}
\label{tab:trec_results}
\renewcommand{\arraystretch}{1.1}
\setlength{\tabcolsep}{4pt}
\footnotesize
\begin{tabular}{l>{\raggedright\arraybackslash}p{1.6cm}cccr}
\toprule
\textbf{Model} & \textbf{Training} & \textbf{Scale (\%)} & \textbf{DL19} $\uparrow$ & \textbf{DL20} $\uparrow$ & \textbf{Avg.} $\uparrow$ \\
\midrule
RankGPT & GPT-4 & - & 75.60 & 70.60 & 73.10 \\
\midrule
RankZephyr-7B & LLM-distil. & 100 & 73.90 & 70.90 & 72.40 \\
Setwise-Rank-R1-7B & SFT & 18 & 73.80 & 69.20 & 71.50 \\
Setwise-Rank-R1-3B & SFT & 18 & 73.40 & 67.20 & 70.30 \\
Setwise-Rank-R1-14B & SFT & 18 & 72.90 & 68.90 & 70.90 \\
Rank-R1-7B & RL & 18 & 72.70 & 68.50 & 70.60 \\
Rank-R1-14B & RL & 18 & 71.40 & 69.10 & 70.25 \\
Rank-R1-3B & RL & 18 & 71.30 & 66.80 & 69.05 \\
Rank-R1-7B & Zeroshot & 100 & 71.20 & 66.20 & 68.70 \\
ListT5-base & SFT & 100 & 71.80 & 68.10 & 69.95 \\
RankVicuna-7b & Zeroshot & 100 & 68.90 & 66.10 & 67.50 \\
Setwise-14B & Zeroshot & 100 & 67.70 & 64.80 & 66.25 \\
Rank-R1-14B & Zeroshot & 100 & 67.90 & 65.20 & 66.55 \\
Setwise-Rank-R1-7B & Zeroshot & 100 & 67.50 & 63.60 & 65.55 \\
RankGPT(GPT-3.5) & GPT-3.5 & - & 65.80 & 62.90 & 64.35 \\
Rank-R1-3B & Zeroshot & 100 & 60.50 & 53.80 & 57.15 \\
BM25 & Zeroshot & 100 & 50.60 & 48.00 & 49.30 \\
Setwise-Rank-R1-3B & Zeroshot & 100 & 37.10 & 31.70 & 34.40 \\
\midrule
\rowcolor{gray!20}
\textbf{Rank-Nexus-2B} & LLM-distil. & \textbf{7.5} & \textbf{74.60} & 70.00 & \textbf{72.30} \\
\bottomrule
\end{tabular}
\vspace{0em}
\end{table}

\textbf{BEIR}~\cite{thakur2021beir} is a heterogeneous zero-shot retrieval benchmark consisting of 18 diverse datasets spanning multiple domains and tasks. We evaluate on 7 representative subsets: COVID, DBPedia, SciFact, NFCorpus, Signal, Robust04, and News. This benchmark tests model generalization to out-of-domain scenarios without task-specific fine-tuning. We follow the standard protocol of reranking BM25-retrieved top-100 passages and report NDCG@10.

\textbf{INQUIRE}~\cite{vendrow2024inquire} is a challenging text-to-image retrieval benchmark featuring expert-level queries on natural world imagery. It includes iNaturalist 2024 (iNat24), a dataset of 5 million natural world images, along with 250 expert-level queries comprehensively labeled with 33,000 total image matches. The queries span categories such as species identification, ecological context, animal behavior, and visual appearance, requiring nuanced image understanding and domain expertise. We report nDCG@50 and MRR as evaluation metrics.

\textbf{MMDocIR}~\cite{dong2025mmdocir} is a multimodal document information retrieval benchmark designed for both page-level and layout-level retrieval tasks. The evaluation set comprises 313 documents (averaging 65.1 pages each) with 1,658 queries and 2,107 page-level annotations plus 2,638 layout-level bounding box labels. The training set contains 73,843 questions from 7 DocQA datasets across 6,878 manually collected documents. This benchmark evaluates the ability to retrieve relevant pages and precise layout elements (paragraphs, figures, tables, charts) from long multimodal documents. We report Recall@1, Recall@3, and Recall@5 for both macro and micro averaged metrics.

\begin{table}[H]
\centering
\setlength{\abovecaptionskip}{0in} 
\setlength{\belowcaptionskip}{0in}
\caption{Results for the INQUIRE-RERANK task on various embedding and multimodal models.}
\label{tab:inquire_results}
\renewcommand{\arraystretch}{1.1}
\setlength{\tabcolsep}{5pt}
\footnotesize
\begin{tabular}{llcc}
\toprule
\textbf{Category} & \textbf{Method} & \textbf{nDCG@50} $\uparrow$ & \textbf{MRR} $\uparrow$ \\
\midrule
Baseline & Random & 52.6 & 0.35 \\
\midrule
\multirow{3}{*}{\textit{Embedding models}} 
& CLIP ViT-B-32 [60] & 59.1 & 0.47 \\
& CLIP ViT-L-14 [60] & 64.2 & 0.57 \\
& CLIP ViT-H-14 [24] & 68.7 & 0.66 \\
\midrule
\multirow{2}{*}{\makecell[l]{\textit{Proprietary}\\\textit{multimodal models}}}
& GPT-4V [3] & 71.9 & 0.70 \\
& GPT-4o [55] & \textbf{78.9} & \textbf{0.78} \\
\midrule
\multirow{3}{*}{\makecell[l]{\textit{Open-source}\\\textit{multimodal models}}}
& LLaVA-v1.6-7B [45] & 70.4 & 0.66 \\
& Qwen3VL-2B & 63.9 & 0.55 \\
& VILA-13B [41] & 71.1 & 0.67 \\
\midrule

\multirow{2}{*}{\textit{Ours}} & \textbf{Rank-Nexus-2B (Qwen)} & 69.0 & 0.60 \\

 & \textbf{Rank-Nexus-2B (InternVL)} & 73.9 & 0.72 \\
\bottomrule
\end{tabular}
\vspace{0em}
\end{table}

\subsection{Implementation Details}

We implement Rank-Nexus using two vision-language model backbones: InternVL-3-2B~\cite{zhu2025internvl3} and Qwen3-VL-2B~\cite{bai2025qwen3vltechnicalreport}, both with 2B parameters to ensure computational efficiency. All experiments are conducted on a single NVIDIA RTX 5090 GPU with 24GB memory. Specific prompt templates used for different stages of our progressive training are detailed in Table~\ref{tab:prompt-engineering}, ensuring consistent instruction following across modalities.

For training, we adopt a parameter-efficient fine-tuning strategy using QLoRA~\cite{dettmers2023qloraefficientfinetuningquantized} with 8-bit quantization. The LoRA configuration uses rank $r=8$ and scaling factor $\alpha=8$, applied to all attention projection matrices. We use a batch size of 4 with gradient accumulation to maintain effective batch sizes while fitting within GPU memory constraints. The learning rate is set to $1\times10^{-5}$ with a cosine annealing schedule and 3\% warmup steps. We train for a maximum of 3 epochs with early stopping based on validation set performance.

For text encoding, we use the pretrained tokenizer from each respective VLM backbone with a maximum sequence length of 512 tokens. Images are resized to the model's default resolution (448×448 for InternVL-3 and 462×462 for Qwen2.5-VL) and normalized using ImageNet statistics. 
\subsection{Main Results}

\subsubsection{Text Reranking Performance}

We evaluate Rank-Nexus on standard text reranking benchmarks and compare it against state-of-the-art methods. Table~\ref{tab:trec_results} presents results on TREC DL19 and DL20 test sets. Despite using only 7.5\% of training data and a compact 2B model, Rank-Nexus achieves 74.60 nDCG@10 on DL19 and 70.00 on DL20, outperforming most existing methods including the 7B RankZephyr model trained on the full dataset. Notably, our model surpasses reasoning-based methods like Rank-R1 while using significantly fewer parameters and requiring no reasoning overhead during inference.

Table~\ref{tab:beir_results} shows zero-shot generalization results on 7 BEIR datasets. Rank-Nexus (InternVL) achieves an average nDCG@10 of 54.92, ranking as the best overall method and outperforming the 14B Rank-R1 model. Particularly strong performance is observed on COVID (85.99), News (53.11), and DBPedia (45.70), demonstrating robust cross-domain transfer capabilities despite being trained on minimal curated data.

\subsubsection{Multimodal Reranking Performance}

On the image-centric INQUIRE benchmark (Table~\ref{tab:inquire_results}), Rank-Nexus achieves 73.9 nDCG@50 and 0.72 MRR, closely approaching GPT-4o (78.9 nDCG@50) while using a 2B open-source model. Our method substantially outperforms other open-source multimodal models including LLaVA-v1.6-7B (70.4) and VILA-13B (71.1), demonstrating effective visual understanding despite the compact model size.

For document-level multimodal retrieval, Table~\ref{tab:mmdocir_results} presents results on the MMDocIR benchmark. We have first introduced retriever-based ranking, which is usually pre-trained encoders to conclude final results by metrics. We chose ColQwen to continue second-stage reranking. When reranking from ColQwen retrievals, Rank-Nexus achieves competitive performance with 70.10 in recall@1, 86.96 in recall@3, showing the model's capability to handle complex multimodal documents containing mixed text, figures, and tables. This validates that our progressive training strategy successfully enables effective joint reasoning over visual and textual content.

\subsection{Ablation Studies}

We conduct comprehensive ablation studies to validate the design choices of Rank-Nexus and understand the contribution of each component.

\subsubsection{Coreset Selection Strategy}

We evaluate the effectiveness of our diversity-based coreset selection compared to random sampling. Table~\ref{tab:ablation_coreset} demonstrates that our diversity-based selection strategy substantially outperforms random sampling when using the same amount of data (1k samples). This validates our approach of prioritizing data quality through systematic diversity-based selection. As the table shows, our method effectively select most useful data that benefits to reranking downstream tasks.

\begin{table}[h]
\centering
\caption{Ablation study on coreset selection strategy. We compare our diversity-based selection against random sampling with equivalent data budgets.}
\label{tab:ablation_coreset}
\small
\begin{tabular}{lcccc}
\toprule
\textbf{Selection Method} & \textbf{Data Scale} & \textbf{DL19} & \textbf{DL20} & \textbf{BEIR Avg} \\
\midrule
Random sampling & 1k & 71.90 & 64.48 & 50.02 \\
\textbf{Diversity-based (Ours)} & 1k & 74.60 & 70.00 & 54.92 \\
K-means Centroid sampling & 1k & 73.23 & 63.47 & 51.10 \\
\bottomrule
\end{tabular}
\end{table}

\begin{table*}[t!]
\centering
\small
\begin{tabular}{|p{0.47\textwidth}|p{0.47\textwidth}|}
\hline
\multicolumn{1}{|c|}{\textbf{Prompt Structure}} & \multicolumn{1}{c|}{\textbf{Example Application}} \\
\hline
\multicolumn{2}{|c|}{\textit{Text-based Listwise Reranking}} \\
\hline
\textbf{System:} "You are RankGPT, an intelligent assistant that can rank passages based on their relevancy to the query." & \multirow{8}{0.47\textwidth}{\textbf{Query:} "What is deep learning?" \newline \textbf{Candidates:} \newline [1] Machine learning is a subset of AI... \newline [2] Deep learning uses neural networks... \newline [3] Python is a programming language... \newline \textbf{Output:} "[2] > [1] > [3]"} \\
\textbf{User:} "I will provide you with \{N\} passages, each indicated by number identifier []. Rank the passages based on their relevance to query: \{query\}." & \\
\textbf{Assistant:} "Okay, please provide the passages." & \\
\textbf{User:} "[\{i\}] \{passage\_i\}" & \\
\textbf{Assistant:} "Received passage [\{i\}]." & \\
\textit{(Repeat for all N passages)} & \\
\textbf{User:} "Search Query: \{query\}. Rank the \{N\} passages above based on their relevance. The output format should be [] > [], e.g., [1] > [2]. Only response the ranking results, do not say any word or explain." & \\
\hline
\multicolumn{2}{|c|}{\textit{Image-based Pairwise Reranking}} \\
\hline
\textbf{System:} "You are an expert relevance assessor for multimodal documents. Determine whether the given document is relevant to the user query." & \multirow{6}{0.47\textwidth}{\textbf{Query:} "Show me annual revenue charts" \newline \textbf{Document:} [Text: "Our Q4 revenue increased 15\%..." + Image: Bar chart showing quarterly revenue] \newline \textbf{Output:} "Yes"} \\
\textbf{User:} "Query: \{query\} \newline Document Text: \{text\} \newline Document Image: [Attached] \newline Is this document relevant to the query? Answer only 'Yes' or 'No'." & \\
\textbf{Expected Response Format:} "Yes" or "No" & \\
\hline
\multicolumn{2}{|c|}{\textit{Image-based Listwise Reranking}} \\
\hline
\textbf{System:} "You are a multimodal reranking assistant. Rank documents containing both text and images based on their relevance to the query." & \multirow{8}{0.47\textwidth}{\textbf{Query:} "Explain transformer architecture" \newline \textbf{Candidates:} \newline [1] Text: "Transformers use attention..." + Image: Architecture diagram \newline [2] Text: "BERT is a language model..." + Image: Training process \newline [3] Text: "CNNs process images..." + Image: Convolution visualization \newline \textbf{Output:} "[1] > [2] > [3]"} \\
\textbf{User:} "I will provide you with \{N\} multimodal documents, each containing text and images. Rank them by relevance to query: \{query\}." & \\
\textbf{Assistant:} "Understood. Please provide the documents." & \\
\textbf{User:} "[\{i\}] Text: \{text\_i\} \newline Image: [Attached image\_i]" & \\
\textbf{Assistant:} "Received document [\{i\}]." & \\
\textit{(Repeat for all N documents)} & \\
\textbf{User:} "Query: \{query\}. Rank the \{N\} documents considering both textual and visual content. Output format: [] > [], e.g., [1] > [2]. Only provide the ranking, no explanation." & \\
\hline
\end{tabular}
\caption{Prompt Engineering Strategies for Text and Multimodal Listwise Reranking}
\vspace{-3mm}
\label{tab:prompt-engineering}
\end{table*}

\section{Discussion}
Our experiments reveal two significant findings regarding the development of multimodal rerankers. First, the data efficiency of Rank-Nexus is remarkable. By leveraging only 7.5\% of the standard training data through diversity-based coreset selection, we achieve performance superior to models trained on the full dataset. This counters the prevailing "scaling law" intuition in retrieval tasks, suggesting that for reranking specifically, the diversity of ranking patterns is more critical than the sheer volume of query-document pairs.

Second, the progressive training curriculum uncovers a strong cross-modal transfer effect. The fact that text-only training improves image retrieval performance suggests that the reasoning logic required for ranking—identifying relevance, comparing attributes, and ordering items—is fundamentally modality-agnostic. The visual encoder acts as a bridge, translating image features into this shared reasoning space. This implies that future multimodal systems can benefit significantly from "warm-starting" with abundant text data before fine-tuning on scarcer multimodal examples.

However, limitations remain. While our 2B model excels in efficiency, it occasionally struggles with extremely fine-grained visual details in complex technical diagrams compared to larger proprietary models like GPT-4o. Additionally, our pipeline currently relies on a specific retriever (ColQwen); future work should investigate end-to-end differentiable retrieval and ranking to further close the modality gap.

\section{Conclusion}
In this paper, we introduced Rank-Nexus, a lightweight 2B parameter multimodal reranker that effectively bridges the gap between text and visual information retrieval. By implementing a novel progressive training strategy and a diversity-based data curation pipeline, we demonstrate that high-performance reranking does not require massive parameter counts or exhaustive datasets. Our approach achieves state-of-the-art results on multiple benchmarks while maintaining high inference speeds suitable for real-world deployment. These findings suggest a promising direction for future research: prioritizing data quality and curriculum learning over raw model scaling to build efficient, robust multimodal information systems.


\bibliographystyle{ACM-Reference-Format}
\bibliography{rank.bib}  


\end{document}